\documentstyle[12pt,epsfig,axodraw]{article}

\advance\textheight by \topskip
\begin{document}
\tolerance=100000
\thispagestyle{empty}
\setcounter{page}{0}

\newcommand{\be}{\begin{equation}}
\newcommand{\ee}{\end{equation}}
\newcommand{\br}{\begin{eqnarray}}
\newcommand{\er}{\end{eqnarray}}
\newcommand{\ba}{\begin{array}}
\newcommand{\ea}{\end{array}}
\newcommand{\bi}{\begin{itemize}}
\newcommand{\ei}{\end{itemize}}
\newcommand{\bn}{\begin{enumerate}}
\newcommand{\en}{\end{enumerate}}
\newcommand{\bc}{\begin{center}}
\newcommand{\ec}{\end{center}}
\newcommand{\ul}{\underline}
\newcommand{\ol}{\overline}
\def\epem{\ifmmode{e^+ e^-} \else{$e^+ e^-$} \fi}
\newcommand{\eeww}{$e^+e^-\rightarrow W^+ W^-$}
\newcommand{\qqQQ}{$q_1\bar q_2 Q_3\bar Q_4$}
\newcommand{\eeqqQQ}{$e^+e^-\rightarrow q_1\bar q_2 Q_3\bar Q_4$}
\newcommand{\eewwqqqq}{$e^+e^-\rightarrow W^+ W^-\ar q\bar q Q\bar Q$}
\newcommand{\eeqqgg}{$e^+e^-\rightarrow q\bar q gg$}
\newcommand{\eeqloop}{$e^+e^-\rightarrow q\bar q gg$ via loop of quarks}
\newcommand{\eeqqqq}{$e^+e^-\rightarrow q\bar q Q\bar Q$}
\newcommand{\eewwjjjj}{$e^+e^-\rightarrow W^+ W^-\rightarrow 4~{\rm{jet}}$}
\newcommand{\eeqqggjjjj}{$e^+e^-\rightarrow q\bar 
q gg\rightarrow 4~{\rm{jet}}$}
\newcommand{\eeqloopjjjj}{$e^+e^-\rightarrow q\bar 
q gg\rightarrow 4~{\rm{jet}}$ via loop of quarks}
\newcommand{\eeqqqqjjjj}{$e^+e^-\rightarrow q\bar q Q\bar Q\rightarrow
4~{\rm{jet}}$}
\newcommand{\eejjjj}{$e^+e^-\rightarrow 4~{\rm{jet}}$}
\newcommand{\jjjj}{$4~{\rm{jet}}$}
\newcommand{\qqbar}{$q\bar q$}
\newcommand{\ww}{$W^+W^-$}
\newcommand{\ar}{\rightarrow}
\newcommand{\sm}{${\cal {SM}}$}
\newcommand{\Dir}{\kern -6.4pt\Big{/}}
\newcommand{\Dirin}{\kern -10.4pt\Big{/}\kern 4.4pt}
\newcommand{\DDir}{\kern -7.6pt\Big{/}}
\newcommand{\DGir}{\kern -6.0pt\Big{/}}
\newcommand{\wwqqqq}{$W^+ W^-\ar q\bar q Q\bar Q$}
\newcommand{\qqgg}{$q\bar q gg$}
\newcommand{\qloop}{$q\bar q gg$ via loop of quarks}
\newcommand{\qqqq}{$q\bar q Q\bar Q$}

\def\st{\sigma_{\mbox{\scriptsize tot}}}
\def\Ord{\buildrel{\scriptscriptstyle <}\over{\scriptscriptstyle\sim}}
\def\OOrd{\buildrel{\scriptscriptstyle >}\over{\scriptscriptstyle\sim}}
\def\pl #1 #2 #3 {{\it Phys.~Lett.} {\bf#1} (#2) #3}
\def\np #1 #2 #3 {{\it Nucl.~Phys.} {\bf#1} (#2) #3}
\def\zp #1 #2 #3 {{\it Z.~Phys.} {\bf#1} (#2) #3}
\def\jp #1 #2 #3 {{\it J.~Phys.} {\bf#1} (#2) #3}
\def\pr #1 #2 #3 {{\it Phys.~Rev.} {\bf#1} (#2) #3}
\def\prep #1 #2 #3 {{\it Phys.~Rep.} {\bf#1} (#2) #3}
\def\prl #1 #2 #3 {{\it Phys.~Rev.~Lett.} {\bf#1} (#2) #3}
\def\mpl #1 #2 #3 {{\it Mod.~Phys.~Lett.} {\bf#1} (#2) #3}
\def\rmp #1 #2 #3 {{\it Rev. Mod. Phys.} {\bf#1} (#2) #3}
\def\cpc #1 #2 #3 {{\it Comp. Phys. Commun.} {\bf#1} (#2) #3}
\def\sjnp #1 #2 #3 {{\it Sov. J. Nucl. Phys.} {\bf#1} (#2) #3}
\def\xx #1 #2 #3 {{\bf#1}, (#2) #3}
\def\hepph #1 {{\tt hep-ph/#1}}
\def\preprint{{\it preprint}}

\begin{flushright}
{\large RAL-TR-97-065}\\
{\large Cavendish-HEP-97/14}\\
{\large LU-TP-97-24}\\ 
%{\rm September 1997\hspace*{.5 truecm}}\\ 
%{\rm Revised November 1997}\\
{\rm November 1997}\\
\end{flushright}

\vspace*{\fill}

\begin{center}
{\Large {\bf $\boldmath{e^+e^-\ar 6}$~jets in parton level QCD\\
at LEP1, LEP2 and NLC\footnote{E-mail:  
Moretti@hep.phy.cam.ac.uk.\\
(*) Address after October 1, 1997: \\
Theoretical Physics Department, Rutherford Appleton Laboratory,
Chilton, Didcot, Oxon OX11 0QX, UK; e-mail: moretti@v2.rl.ac.uk}}}\\[1.cm]
{\large 
S.~Moretti$^{(*)}$}\\[0.4 cm]
{\it Cavendish Laboratory, University of Cambridge,}\\
{\it Madingley Road, Cambridge CB3 0HE, UK.}
\end{center}

\vspace*{\fill}

\begin{abstract}
{\normalsize
\noindent
We study electron-positron annihilations into six jets at the parton level in 
perturbative Quantum Chromo-Dynamics. We use helicity amplitude methods.
Results are presented for the case of the Durham and Cambridge jet
clustering algorithms at three different collider energies.  
}
\end{abstract}

\vspace*{\fill}
\newpage

\section*{1. Introduction and motivation}

Experimental collaborations at LEP have studied the physics of hadronic
jets in great detail. The list of references is huge and we 
recommend the reader to look at any of the various
high energy physics databases and exploit the appropriate keyword search, 
rather then trying to reproduce them here. What we would like to point
out is that the phenomenology of six-jet events is not at all well known to
date, from both the experimental and theoretical point of view. 
Indeed, this is not surprising, since
the complications arising from, on the one hand,
the poor event rate and large number of tracks, 
and, on the other hand, the complexity of the 
perturbative QCD calculations, are difficult to overcome.

Nonetheless, it is important that we soon examine
this issue. Why ? Well, we could stand on the fact that ALEPH sees 
at 161 GeV a large excess of five- and especially six-jets events \cite{6j}
(see also \cite{GC}) and speculate about its significance and the need
for reliable perturbative calculations.
We rather prefer to make a more far-seeing consideration. As for hadronic
physics in electron-positron annihilations we can say the following. 
LEP1 was the era of the $Z$-peak and of its two-jet
(and several higher order) decays. LEP2 is the age of the $W^+W^-$-resonance 
and of its four-jet (and some higher order) decays. 
As we will move into the NLC epoch, we will step into
a long series of resonant processes ending up with six-jet signatures,
typically. One can mention top physics for a start. As this is one
of the main goals of NLC, we should expect a lot of experimental
studies concerned with $t\bar t\ar b\bar b W^+W^-\ar \mbox{6~jet}$ events,
as the $W^\pm$'s show the `colourful' tendency of decaying hadronically, though
one would rather prefer to exploit a mixed semileptonic signature 
(in which one of the $W^\pm$'s accomplishes an electron or muon decay), to
make things easier in terms of multi-jet resolution and mass reconstruction. 
Then one should not 
forget the new generation of gauge boson resonances, such as $W^+W^-Z$ and
$ZZZ$,
and their favorite decays, needless to say, into 
six jets\footnote{One could also add photonic processes, like $\gamma W^+W^-$, 
$\gamma ZZ$, $\gamma\gamma Z$ and $\gamma\gamma\gamma$ (though the photons 
would not be so virtual to easily comply with the tendency of eventually
yielding a pair of partons) as well as those involving the Higgs particle, via 
$ZH\ar ZW^+W^-$ and $ZH\ar ZZZ$ to detect it and via 
$ZH\ar ZHH$ to measure its self interactions.}. 
Indeed, for accurate studies of these kinds of events, it is worthwhile
to use all the decay signatures, not only those involving leptons. 

It is not our task here to remind the reader why top-antitop and 
three-gauge-boson production are relevant in the future of particle physics.
To this end, one can find a complete compilation of motivations,
procedures and expectations in Ref.~\cite{ee500}. What we would like
to stress here
is that one should be ready with all the appropriate phenomenological
instruments to challenge the new kind of multi-jet experimental studies
that will have to be carried out at the new generation of $e^+e^-$ machines.

As for the theoretical progress in this respect, 
studies of $e^+e^-\ar $ six fermion electroweak processes are under way. 
A brief account of
approaches and methods can be found in Ref.~\cite{nico5}. These kinds of
reactions have already been calculated and analysed for the case involving up 
to four-quarks in the final state \cite{sandro} (and other similar studies
are in preparation 
\cite{grace}), while in Ref.~\cite{nico} the concern was about
Higgs processes with two quarks produced.
One should probably expect the upgrade of the codes used for those
studies to the case of six-quark production via electroweak interactions
quite soon. However, a large fraction 
of the six-jet cross section  comes
from QCD interaction involving gluon propagators, gluon emissions and 
quark-gluon couplings. 

In the present paper, we by-pass the case of ${\cal O}(\alpha_s^2)$ 
interactions. Those involving two quarks
represent in fact a trivial extension of the projects carried out in 
Ref.~\cite{sandro}--\cite{nico}. Those involving two gluons have been dealt
with for the case of $W^+W^-\ar6$~jet decays in Ref.~\cite{WW6} and it would  
be straightforward to extend those calculations to the case of 
$\gamma\gamma,\gamma Z,ZZ\ar6$~jet decays as well. As for the other channels, 
they will be discussed elsewhere \cite{prep}. 
Here, we address the case
of six-jet production through  ${\cal O}(\alpha_s^4)$ in perturbative
QCD. That is, we calculate the tree-level processes 
\be
\label{6j}
e^+e^-\ar q\bar q gggg, 
\qquad
e^+e^-\ar q\bar q q'\bar q' gg, 
\qquad
e^+e^-\ar q\bar q q'\bar q' q''\bar q'',
\ee
where $q,q',q''$ represent {\sl massless} quarks and $g$ a gluon. We will apply
our results to the case of LEP1, LEP2 and NLC energies. A short description
of the computational techniques adopted will be given in Section 2.
Results and conclusions are in Section 3.

\section*{2. Matrix Elements}

\begin{figure}[!t]
\begin{center}
\vskip-2.0cm
\begin{picture}(500,100)
\SetScale{.9}
\SetWidth{1.2}
\SetOffset(0,0)

\Text(30,110)[]{(a)}

\Text(90,63.25)[]{$\times$}
\Line( 50,70)(150,70)
\Gluon(70,70)(70,115){3}{3}
\Gluon(90,70)(90,115){3}{3}
\Gluon(110,70)(110,115){3}{3}
\Gluon(130,70)(130,115){3}{3}

\Text(225,63.25)[]{$\times$}
\Line(200,70)(300,70)
\Gluon(220,70)(220,115){3}{3}
\Gluon(240,70)(240,115){3}{3}
\Gluon(270,70)(270,95){3}{2}
\Gluon(270,95)(255,115){3}{2}
\Gluon(285,115)(270,95){3}{2}

\Text(360,63.25)[]{$\times$}
\Line(350,70)(450,70)
\Gluon(365,70)(365,115){3}{3}
\Gluon(415,70)(415,95){3}{2}
\Gluon(415,95)(415,115){3}{2}
\Gluon(415,95)(385,115){3}{3}
\Gluon(445,115)(415,95){3}{3}

\Text(90,-4.5)[]{$\times$}
\Line( 50,-5)(150,-5)
\Gluon(65,-5)(65,40){3}{3}
\Gluon(90,-5)(90,20){3}{2}
\Gluon(90,20)(90,40){3}{2}
\Gluon(90,20)(115,20){3}{2}
\Gluon(115,20)(115,40){3}{2}
\Gluon(115,20)(140,20){3}{2}

\Text(225,-4.5)[]{$\times$}
\Line(200,-5)(300,-5)
\Gluon(230,-5)(230,20){3}{2}
\Gluon(230,20)(215,40){3}{2}
\Gluon(245,40)(230,20){3}{2}
\Gluon(270,-5)(270,20){3}{2}
\Gluon(270,20)(255,40){3}{2}
\Gluon(285,40)(270,20){3}{2}

\Text(360,-4.5)[]{$\times$}
\Line(350,-5)(450,-5)
\Gluon(375,-5)(375,20){3}{2}
\Gluon(375,20)(375,40){3}{2}
\Gluon(375,20)(410,20){3}{3}
\Gluon(410,20)(410,40){3}{2}
\Gluon(430,20)(410,20){3}{2}
\Gluon(410,0)(410,20){3}{2}

\Text(90,-72)[]{$\times$}
\Line(50,-80)(150,-80)
\Gluon(65,-80)(65,-55){3}{2}
\Gluon(65,-55)(65,-35){3}{2}
\Gluon(65,-55)(90,-55){3}{2}
\Gluon(90,-55)(90,-35){3}{2}
\Gluon(90,-55)(115,-55){3}{2}
\Gluon(115,-55)(115,-35){3}{2}
\Gluon(115,-55)(140,-55){3}{2}

\Text(225,-72)[]{$\times$}
\Line(200,-80)(300,-80)
\Gluon(235,-80)(235,-55){3}{2}
\Gluon(235,-55)(235,-35){3}{2}
\Gluon(235,-55)(215,-55){3}{2}
\Gluon(235,-55)(260,-55){3}{3}
\Gluon(260,-55)(260,-35){3}{2}
\Gluon(260,-55)(280,-55){3}{2}

\end{picture}
%\vskip3.0cm
%\caption{Feynman topologies contributing in lowest order to $e^+e^-\ar 
%q\bar q gggg$. The symbol $\times$ refers to the insertion of the $e^+e^-\ar
%\gamma,Z$ current. Permutations are not shown.} 
%\label{fig_2q4g}
\end{center}
\begin{center}
\vskip4.0cm
\begin{picture}(500,170)
\SetScale{.9}
\SetWidth{1.2}
\SetOffset(0,0)

\Text(30,180)[]{(b)}

\Text(225,131)[]{$\times$}
\Line(200,145)(300,145)
\Gluon(225,145)(225,165){3}{2}
\Gluon(225,165)(205,190){3}{2}
\Gluon(245,190)(225,165){3}{2}
\Gluon(275,145)(275,170){3}{2}
\Line(255,190)(275,170)
\Line(295,190)(275,170)

\Text(90,63.25)[]{$\times$}
\Line( 50,70)(150,70)
\Gluon(70,70)(70,115){3}{3}
\Gluon(90,70)(90,115){3}{3}
\Gluon(125,70)(125,95){3}{2}
\Line(105,115)(125,95)
\Line(145,115)(125,95)

\Text(225,63.25)[]{$\times$}
\Line(200,70)(300,70)
\Gluon(220,70)(220,115){3}{3}
\Gluon(260,70)(260,95){3}{2}
\Line(240,115)(260,95)
\Line(280,115)(260,95)
\Gluon(270,105)(260,115){3}{2}

\Text(360,63.25)[]{$\times$}
\Line(350,70)(450,70)
\Gluon(370,70)(370,115){3}{3}
\Gluon(410,70)(410,87.5){3}{2}
\Gluon(410,87.5)(410,105){3}{2}
\Line(390,115)(410,105)
\Line(430,115)(410,105)
\Gluon(430,87.5)(410,87.5){3}{2}

\Text(90,-4.5)[]{$\times$}
\Line( 50,-5)(150,-5)
\Gluon(90,-5)(90,10){3}{2}
\Line(60,40)(90,10)
\Line(120,40)(90,10)
\Gluon(75,25)(85,40){3}{2}
\Gluon(95,40)(105,25){3}{2}

\Text(225,-4.5)[]{$\times$}
\Line(200,-5)(300,-5)
\Gluon(240,-5)(240,10){3}{2}
\Line(210,40)(240,10)
\Line(270,40)(240,10)
\Gluon(235,25)(245,15){3}{2}
\Gluon(220,40)(235,25){3}{2}
\Gluon(235,25)(250,40){3}{2}

\Text(360,-4.5)[]{$\times$}
\Line(350,-5)(450,-5)
\Gluon(390,-5)(390,10){3}{2}
\Gluon(390,10)(390,25){3}{2}
\Gluon(390,10)(410,10){3}{2}
\Gluon(390,10)(370,10){3}{2}
\Line(370,40)(390,25)
\Line(410,40)(390,25)

\Text(90,-72)[]{$\times$}
\Line(50,-80)(150,-80)
\Gluon(90,-80)(90,-66){3}{2}
\Gluon(90,-66)(70,-66){3}{2}
\Gluon(90,-66)(90,-52){3}{2}
\Gluon(110,-52)(90,-52){3}{2}
\Gluon(90,-52)(90,-38){3}{2}
\Line(70,-35)(90,-38)
\Line(110,-35)(90,-38)

\Text(225,-72)[]{$\times$}
\Line(200,-80)(300,-80)
\Gluon(240,-80)(240,-59){3}{3}
\Gluon(240,-59)(240,-38){3}{3}
\Gluon(240,-59)(270,-59){3}{3}
\Gluon(270,-59)(270,-42){3}{2}
\Gluon(270,-76)(270,-59){3}{2}
\Line(220,-35)(240,-38)
\Line(260,-35)(240,-38)

\Text(360,-72)[]{$\times$}
\Line(350,-80)(450,-80)
\Gluon(390,-80)(390,-65){3}{2}
\Gluon(390,-65)(390,-50){3}{2}
\Gluon(390,-65)(410,-65){3}{2}
\Line(370,-35)(390,-50)
\Line(410,-35)(390,-50)
\Gluon(397,-45)(382,-35){3}{2}

\end{picture}
%\vskip2.5cm
%\caption{Feynman topologies contributing in lowest order to $e^+e^-\ar 
%q\bar q q'\bar q' gg$. The symbol $\times$ refers to the insertion of 
%the $e^+e^-\ar \gamma,Z$ current. Permutations are not shown.} 
%\label{fig_4q2g}
\end{center}
\begin{center}
\vskip4.25cm
\begin{picture}(500,100)
\SetScale{.9}
\SetWidth{1.2}
\SetOffset(0,0)

\Text(30,110)[]{(c)}

\Text(90,63.25)[]{$\times$}
\Line(50,70)(150,70)
\Gluon(90,70)(90,85){3}{2}
\Line(60,115)(90,85)
\Line(120,115)(90,85)
\Gluon(85,100)(95,90){3}{2}
\Line(70,115)(85,100)
\Line(85,100)(100,115)

\Text(225,63.25)[]{$\times$}
\Line(200,70)(300,70)
\Gluon(275,70)(275,95){3}{2}
\Line(255,115)(275,95)
\Line(295,115)(275,95)
\Gluon(225,70)(225,95){3}{2}
\Line(205,115)(225,95)
\Line(245,115)(225,95)

\Text(360,63.25)[]{$\times$}
\Line(350,70)(450,70)
\Gluon(390,70)(390,91){3}{3}
\Gluon(390,91)(390,112){3}{3}
\Gluon(390,91)(420,91){3}{3}
\Line(420,91)(425,108)
\Line(425,74)(420,91)
\Line(370,115)(390,112)
\Line(410,115)(390,112)

\end{picture}
\vskip-0.5cm
\caption{Tree-level Feynman topologies contributing to:
(a) $e^+e^-\ar q\bar q gggg$, 
(b) $e^+e^-\ar q\bar q q'\bar q' gg$ and
(c) $e^+e^-\ar q\bar q q'\bar q' q''\bar q''$. 
The symbol $\times$ refers to the insertion of 
the $e^+e^-\ar \gamma,Z$ current. Permutations are not shown.} 
%\label{fig_6q}
\label{fig_6partons}
\end{center}
\end{figure}
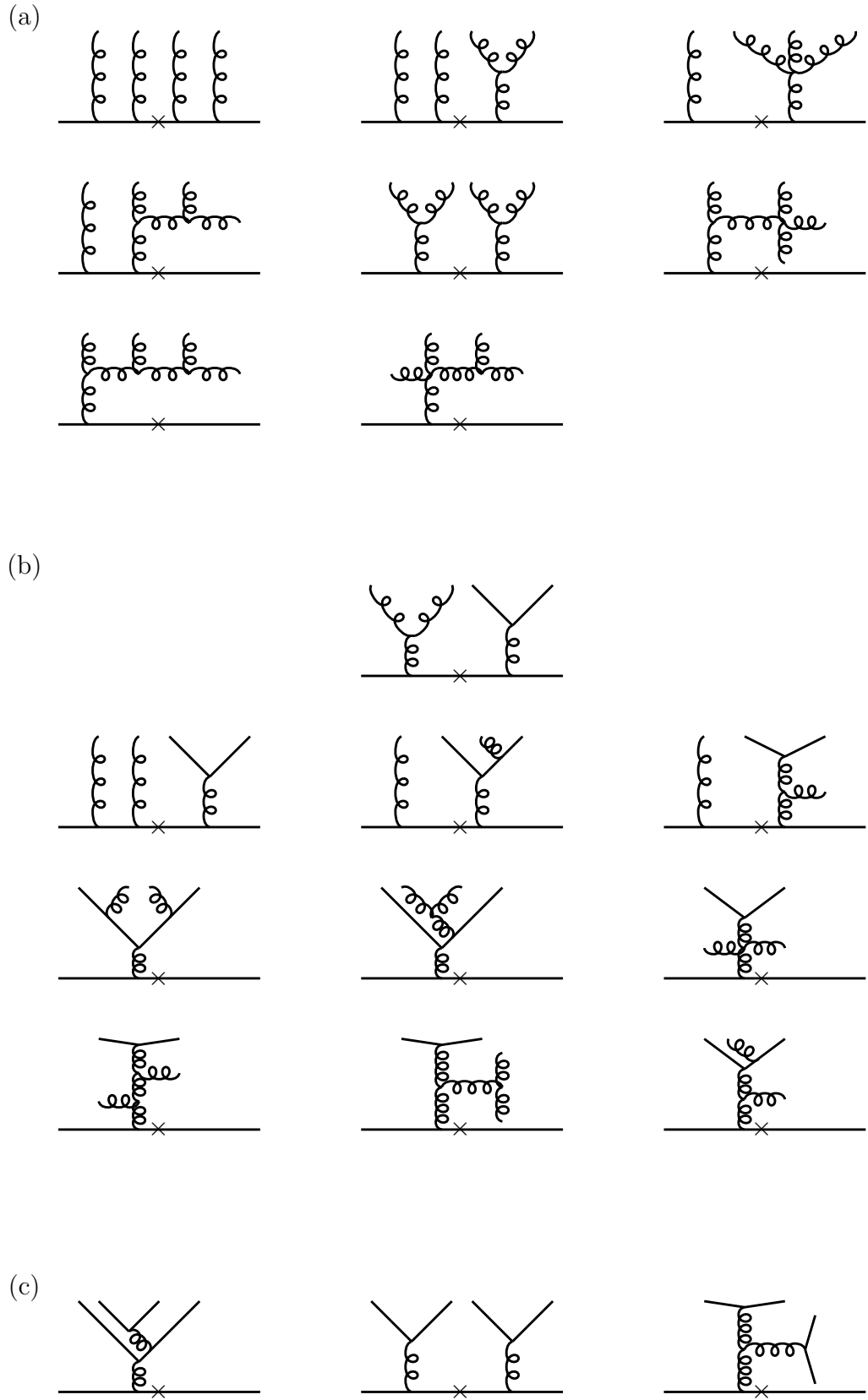

In order to master the large number of Feynman diagrams (of the order
of several hundreds) entering in 
processes (\ref{6j}), we have used spinor techniques \cite{HZ}
and exploited the {\tt HELAS} subroutines \cite{HELAS}. The coded
helicity amplitudes have been checked for gauge invariance and integrated
using {\tt VEGAS} \cite{VEGAS}. 
The three colour matrices have been calculated using the orthogonal basis 
method of Ref.~\cite{color}, which reduces considerably the difficulty of such
computation. The expressions of both the amplitudes and the colour
factors will be given elsewhere \cite{prep}. In the present publication, we 
only mention that the number of different `topologies'
describing processes (\ref{6j}) is rather small in the end, and the
same implementation can be exploited several times by means of recursive 
permutations of the momenta of the external particles. Furthermore, each
common `sub-diagram' is saved and then reused in the same numerical evaluation.
In the case of two-quark-four-gluon diagrams the topologies are those shown in 
Fig.~\ref{fig_6partons}(a), eight in total. 
For four-quark-two-gluons, Fig.~\ref{fig_6partons}(b), one has ten
and for six-quarks, Fig.~\ref{fig_6partons}(c), three topologies.

\vfill\clearpage

\vfill\clearpage

We have adopted $M_Z=91.17$ GeV, $\Gamma_Z=2.516$ GeV,
$\sin^2 (\theta_W)=0.23$, $\alpha_{em}= 1/128$ and the two-loop expression
for $\alpha_{s}$.
As centre-of-mass (CM) energies representative of LEP1, LEP2 and NLC, we have
used the values $\sqrt s=M_Z,180$ GeV and 500 GeV, respectively.

As jet clustering algorithms we have adopted the Durham (D) \cite{durham}
and Cambridge (C) \cite{cambridge} schemes, which are based on the same
`jet measure' 
\be
\label{measure}
y_{ij} = {{2\min (E^2_i, E^2_j)(1-\cos\theta_{ij})}
\over{s}},
\ee
but differ in the clustering procedure. In eq.~(\ref{measure}), 
$E_i$ and $E_j$ represent the energies of any pair of partons $i$ 
and $j$, whereas $\theta_{ij}$ is their relative angle. A six-jet sample
is selected by imposing the constraint $y_{ij}\ge y$ on all $n$ possible
parton combinations $ij$ (with $n=15$ for the Durham and $5\leq n\leq 15$
for the Cambridge algorithm).

\section*{3. Results}

\begin{figure}[htb]
\begin{center}
~\epsfig{file=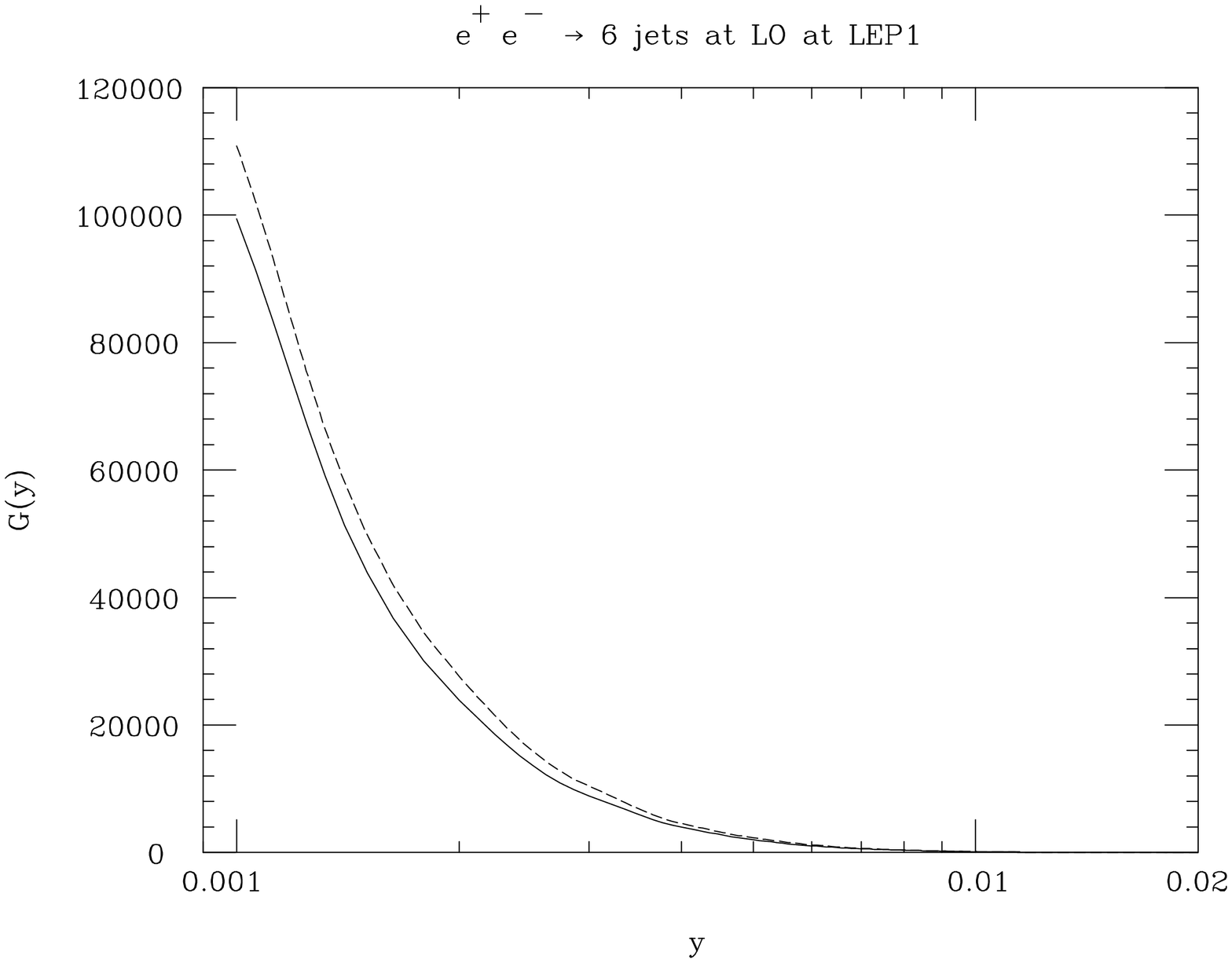,width=16cm,height=12cm,angle=90}
\caption{The parton level $G(y)$ function entering in the six-jet fraction
at LO in the D (continuous line) and C (dashed line) schemes at LEP1.}
\label{fig_jet6}
\end{center}
\end{figure}

We define the $y$-dependent six-jet fraction\footnote{The jet resolution 
parameter $y$ is often indicated as $y_{\rm{\tiny{cut}}}$ 
in the specialised literature.} by means of the relation
\be
\label{def6}
f_6(y)=\frac{\sigma_6(y)}
                  {\sum_m \sigma_m(y)}
            =\frac{\sigma_6(y)}
                  {\st},
\ee
where $\sigma_6(y)$ is the actual six-parton cross section and $\st$
identifies the {\sl total} hadronic rate
$\st=\sigma_{0}(1+\alpha_s/\pi+ ... )$, $\sigma_{0}$ being the 
Born cross section. In perturbative QCD one can rewrite eq.~(\ref{def6})
in terms of a series in $\alpha_s$, beginning with its fourth power, as
\be
\label{f6}
f_6(y) =     \left( \frac{\alpha_s}{2\pi} \right)^4  G(y) + ... , 
\ee
where $G(y)$ is the lowest-order (or leading-order, LO)
`coefficient function' of the six-jet rate.
We show this quantity in Fig.~\ref{fig_jet6}, for the case of both 
the Durham and Cambridge schemes, at LEP1\footnote{Here and in the 
following, unless otherwise stated, 
the summations over the three reactions (\ref{6j}) and over all possible 
combinations of quark flavours in each of these have been performed.}.

The larger rate for the C scheme as compared to the D one is
a direct consequence of the `soft freezing' procedure of resolved jets
described in  Ref.~\cite{cambridge}. The step of eliminating
from the sequence of clustering 
the less energetic one in a resolved pair of particles (i.e., 
with $y_{ij}>y$), implemented in the C algorithm,
tends to enhance the final jet multiplicity
of the original D scheme. In fact, the procedure prevents the attraction 
of the remaining particles (at large angle) into new unresolved pairs,
whose momenta would then be merged together, producing a lower number of 
final jets. For example, at $y=(0.001)[0.005]\{0.010\}$ relative differences
between the two algorithms are of the order of $(10)[14]\{16\}\%$.

The jet fraction $f_6(y)$ and cross section $\sigma_6(y)$  
corresponding to the rates in Fig.~\ref{fig_jet6} 
are given in Tab.~\ref{tab_jet6}, for a representative selection
of $y$'s. The value of $\alpha_s$ adopted is 0.120, whereas that
used for $\st$ is 39.86 nb. 
Given the total luminosity collected at LEP1 in the 1989-1995 years (more than
$10^7$ hadronic events have been recorded by the four collaborations), the 
six-jet event rate is comfortably measurable. For example, for a luminosity
of, say, 100 pb$^{-1}$ per experiment, perturbative QCD predicts at LO
some 52,000 original six-parton events recognised as six-jet ones, for
$y=0.001$ in the D scheme (a number that increases to approximately 58,000
if the C algorithm is adopted instead). Indeed, six-jet fractions have been 
studied at LEP1 in several experimental \cite{jetfractionsLEP1}
and theoretical papers \cite{garth}.

\begin{table}[t]
\begin{center}
\begin{tabular}{|c|c|c|}
\hline
\rule[0cm]{0cm}{0cm}
$y$ & $f_6(y)$ & $\sigma_6(y)$ (pb) \\ \hline\hline
\rule[0cm]{0cm}{0cm}
$0.001$ & $(1.31)[1.47]\times10^{-2}$  & $(523.91)[584.19]$ \\
$0.002$ & $(3.16)[3.65]\times10^{-3}$  & $(125.95)[145.44]$ \\
$0.003$ & $(1.17)[1.37]\times10^{-3}$  & $(46.81)[54.72]$ \\
$0.004$ & $(5.29)[6.01]\times10^{-4}$  & $(21.09)[23.96]$ \\
$0.005$ & $(2.63)[3.05]\times10^{-4}$  & $(10.49)[12.17]$ \\
$0.006$ & $(1.38)[1.59]\times10^{-4}$  & $(5.49)[6.35]$ \\
$0.007$ & $(7.90)[8.85]\times10^{-5}$  & $(3.15)[3.53]$ \\
$0.008$ & $(4.80)[5.20]\times10^{-5}$  & $(1.91)[2.07]$ \\
$0.009$ & $(2.77)[3.08]\times10^{-5}$  & $(1.10)[1.23]$ \\
$0.010$ & $(1.64)[1.93]\times10^{-5}$  & $(0.65)[0.77]$ \\ \hline\hline
\multicolumn{3}{|c|}
{\rule[0cm]{0cm}{0cm}
LEP1} \\ \hline
\end{tabular}
\caption{Jet fraction and cross section rates for $e^+e^-\ar$~6~jets 
at LEP1, for several values of $y$ in the Durham and Cambridge schemes,
in round and squared brackets, respectively. The numerical errors do not
affect the significative digits shown.}
\label{tab_jet6}
\end{center}
\end{table}
 
The rates given in Fig.~\ref{fig_jet6} 
and Tab.~\ref{tab_jet6} are LO results. 
Next-to-leading order (NLO) corrections proportional to ${\cal O}(\alpha_s^5)$
are expected to be large. In fact, the size of higher order (HO) 
contributions generally increases with the
power of $\alpha_s$ and with the number
of particles in the final state or, in other terms, with 
their possible permutations (that is, with the number of different
possible attachments of both the additional 
real and virtual particles to those appearing in lowest-order). 
The highest-order corrections calculated to date in $e^+e^-\ar n$~jet
annihilations are the NLO ones to the four-jet rate, that is,  terms
proportional to $\alpha_s^3$ \cite{slac}--\cite{budapest}. The total 
four-jet cross section at NLO was found to be larger
than that obtained at LO in $\alpha_s^2$ by a factor 1.5 or more, depending
on the value of $y$ implemented during the clustering procedure and
the algorithm used as well. 
Therefore, we should expect that the rates given in Tab.~\ref{tab_jet6}
underestimate the six-jet rates by {\sl at least} a similar factor.

In this respect the Monte Carlo (MC) programs \cite{Jetset}--\cite{Ariadne} 
largely exploited in experimental analyses perform better than the fixed
order perturbative calculations, especially at low $y$-values, where
the latter need to be supported by the additional contribution of
perturbation series involving leading and next-to-leading powers of $\log y$ 
resummed to all orders (for the case of the D scheme, see 
Refs.~\cite{CDOTW,BS2}) in order to fit the same data.
However, as such MCs generally implement
only the infrared (i.e., soft and collinear) dynamics of 
quarks and gluons in the standard `parton-shower (PS) $+ {\cal O}(\alpha_s)$
Matrix Element (ME)' modeling, in many cases their description of the 
large $y$-behaviour and/or that of the interactions of secondary
`branching products' is (or should be expected to be) no longer adequate. 
In fact, this has been shown to be the case, e.g., for some typical
angular quantities of four-jet events \cite{GC,ALEPHgluino}. In
contrast, once ${\cal O}(\alpha_s^2)$ MEs are inserted and properly matched
to the parton shower, e.g., using the JETSET string fragmentation model
\cite{Pythia} (see also Ref.~\cite{andre})\footnote{A special
`${\cal O}(\alpha_s^2) + \mbox{PS} + \mbox{cluster hadronisation}$' 
version of HERWIG is also in preparation.}, then the agreement is 
recovered. In this context, one should however mention that  
ME models with `added-on' hadronisation cannot be reliably extrapolated
from one energy to another, as the fragmentation tuning is 
energy dependent. If one considers that four-jet events represent only the 
next-to-lowest order QCD interactions in $e^+e^-$ scatterings, then
it is not unreasonable to argue that further complications might well
arise as the final state considered gets more and more sophisticated, such as in
five- \cite{five}
and six-jet events. Under these circumstances, we believe the
availability of exact perturbative calculations of the latter to be essential 
to model the HO parton dynamics for both future tests of QCD 
and QCD background studies as well.

\begin{figure}[htb]
\begin{center}
~\epsfig{file=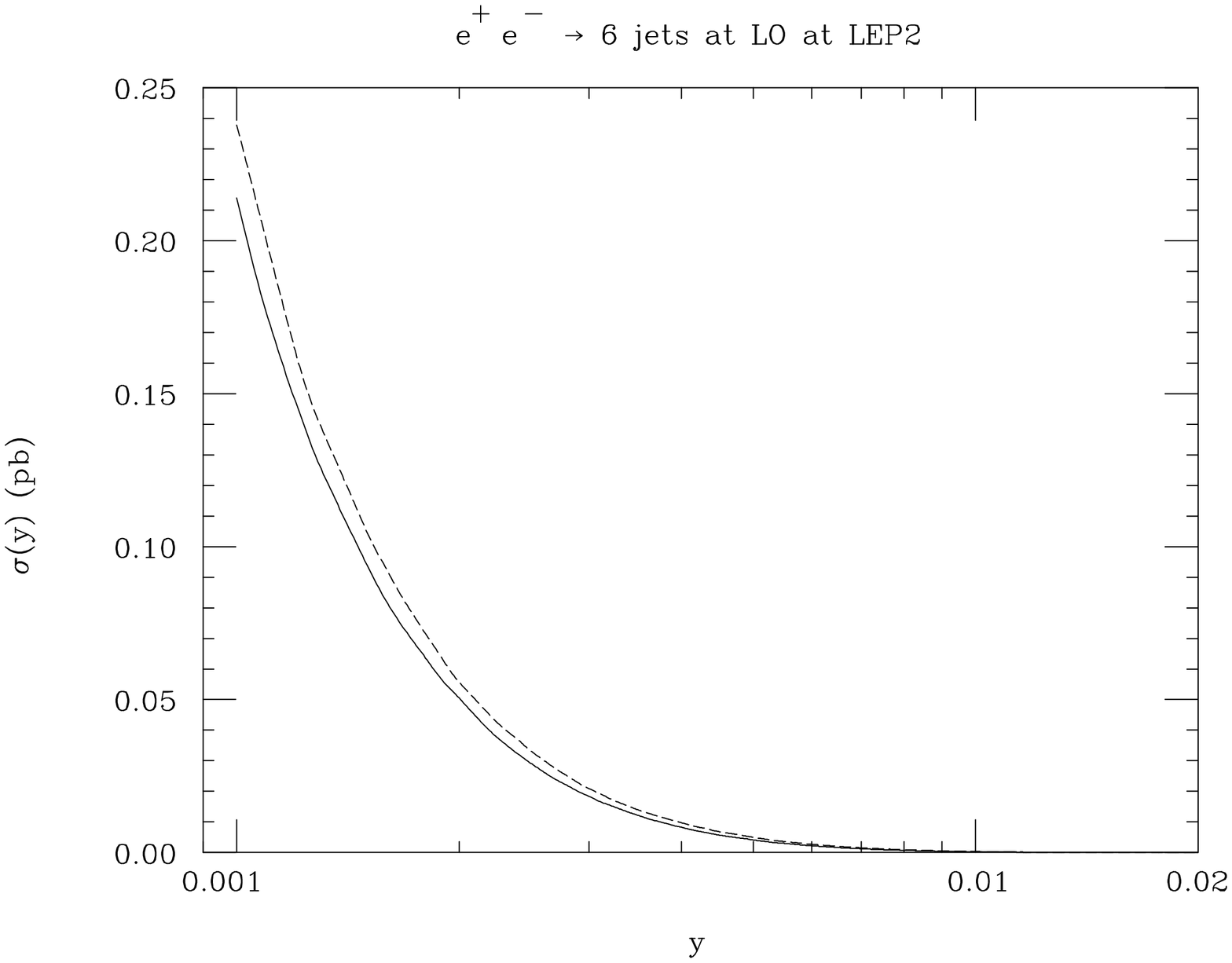,width=16cm,height=12cm,angle=90}
\caption{The total cross section of six-jet events
at LO in the D (continuous line) and C (dashed line) schemes at LEP2.}
\label{fig_jet6lep2}
\end{center}
\end{figure}

The total hadronic cross section falls drastically 
when increasing the CM energy from
the LEP1 to the LEP2 values, by more than three orders of magnitude,
and so does the six-jet rate. Nonetheless, six-jet fractions have been
measured also during the 1995-1996 runs at the CM energies of 130--136,
161 and 172 GeV, when a total luminosity of around 27 pb$^{-1}$ was collected.
Results can be found in Ref.~\cite{jetfractionsLEP2}.
Within a `typical' K-factor of 2 
(which quantifies the ratio between the NLO and the LO six-jet rates)
the values of $f_6(y)$ as reconstructed from our rates  
are always well compatible with those produced by the MCs 
(also at $\sqrt s=161$ GeV, where the six-jet excess was observed) used by, 
e.g., the ALEPH collaboration \cite{6j}. For reference, we present
the jet rates at $\sqrt s=180$ GeV in Fig.~\ref{fig_jet6lep2} in the
form of the total cross section. The relative differences between
the two D and C algorithms are similar to those already
encountered at LEP1.

The six-jet event rate that will be collected at
 NLC can be rather large, despite the cross section being more than a factor
$10^4$ smaller than at LEP1 (e.g., at $\sqrt s=500$ GeV).
This is due to the large yearly luminosity expected 
at this machine, around 10 fb$^{-1}$. For such a value and assuming a 
standard evolution of the coupling constant with the increasing energy 
(that is, no non-Standard Model thresholds occur up to 500 GeV), at the minimum
of the $y$-values considered here (i.e., $y=0.001$), one should expect some  
220 events per annum by adopting the D scheme and about 13\% more if one
adopts the C one. However, these rates decrease rapidly as the resolution 
parameter gets larger, by a factor of 50 or so at $y=0.005$ and of 
approximately 800 at $y=0.01$. 
 
\begin{figure}[tbh]
\begin{center}
~\epsfig{file=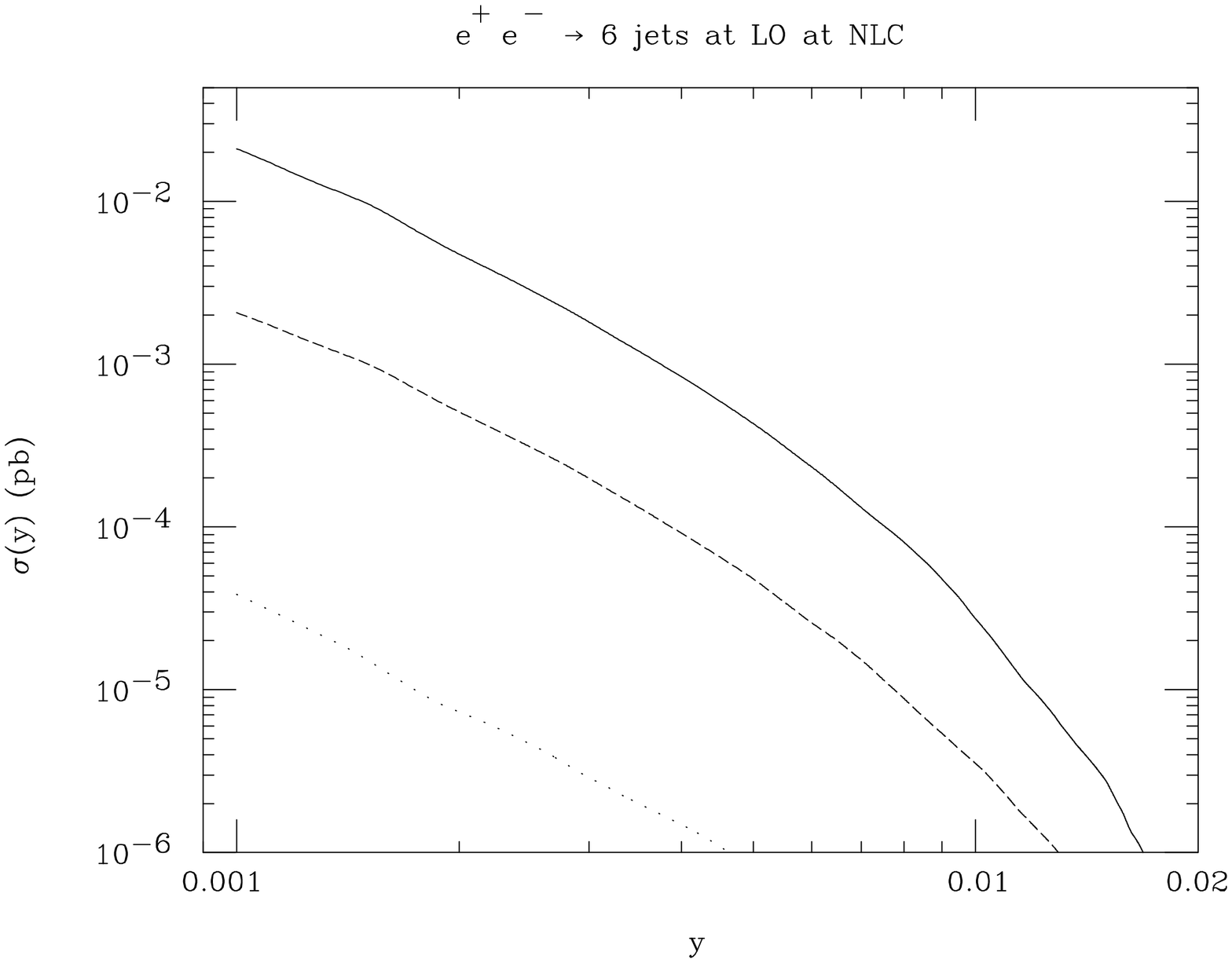,width=16cm,height=12cm,angle=90}
\caption{The total cross section of six-jet events 
at LO in the C scheme at NLC decomposed in terms of the three contributions
$e^+e^-\ar q\bar q gggg$ (continuous line), $e^+e^-\ar q\bar q q'\bar q' gg$
(dashed line) and $e^+e^-\ar q\bar q q'\bar q' q''\bar q''$ (dotted line).}
\label{fig_comp6nlc}
\end{center}
\end{figure}

It is also interesting to look at the composition of the 
total rates in terms of the three processes (\ref{6j}). Whereas this is
probably of little concern at LEP1 and LEP2, the capability of the detectors
of distinguishing between jets due to quarks (and among these, bottom ones
in particular: e.g., in tagging top and/or Higgs decays) 
and gluons is essential at NLC, in order to
perform dedicated searches for signals of both anomalous gauge couplings 
and new particles. 
The different behaviours of the three reactions in (\ref{6j})
can be appreciated in Fig.~\ref{fig_comp6nlc}, e.g., 
for the case of the C scheme, in terms of total cross sections.
The rates for the D algorithm follow
the same pattern. 
%Note that the relative 
%cross sections among the three parton reactions would not be significantly
%different at LEP2 whereas at LEP1 the four-quark-two-gluon and six-quark
%rates are larger as compared to the two-quark-four-gluon one, with a
%proportion (at the lower end of the $y$-spectrum) of the sort 
%$\sigma(q\bar q gggg):\sigma(q\bar q q'\bar q'gg):\sigma(q\bar q
%q'\bar q' q''\bar q'')\simeq 192:46:1$.

\begin{figure}[tbh]
\begin{center}
~\epsfig{file=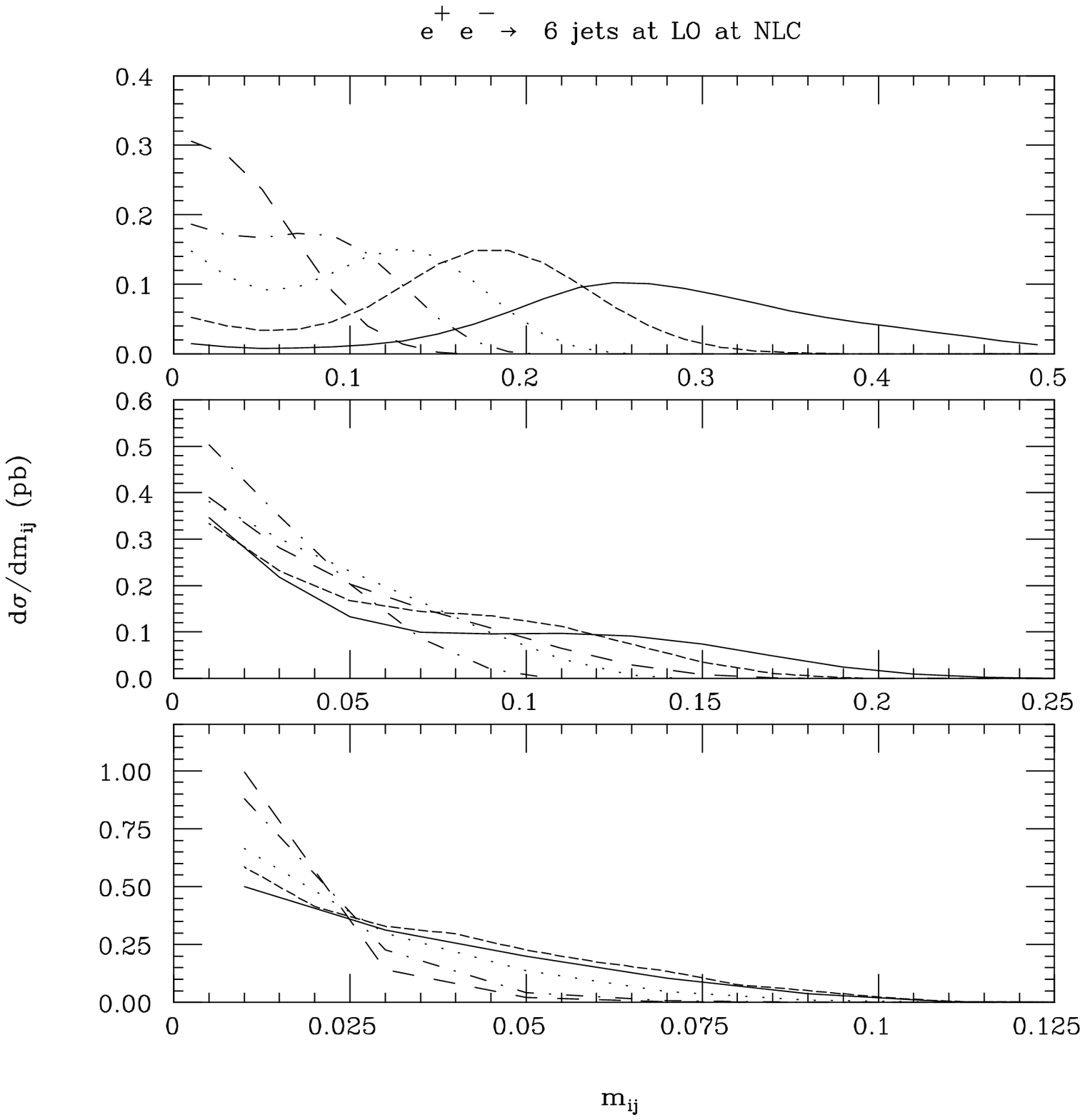,width=16cm,height=18cm,angle=0}
\caption{The distributions in the reduced invariant mass $m_{ij}$ (\ref{mij}) 
of events of the type (\ref{6j}), in the C scheme
with $y=0.001$, for the following combinations of parton pairs $ij$: 
$({12})[{23}]\{{35}\}$ (continuous lines),  
$({13})[{24}]\{{36}\}$ (short-dashed lines),  
$({14})[{25}]\{{45}\}$ (dotted lines),  
$({15})[{26}]\{{46}\}$ (dot-dashed lines) and  
$({16})[{34}]\{{56}\}$ (long-dashed lines),  
in the (upper)[central]\{lower\} frame.}
\label{fig_massesnlc}
\end{center}
\end{figure}

Tests of multiple electroweak self-couplings of  gauge-bosons  
(as well as searches for new resonances) will often need to 
rely on the mass reconstruction of multi-jet systems (particularly
di-jet ones). Therefore, it is 
instructive to look at the invariant mass distributions
which will be produced at NLC 
by all the possible two-parton combinations $ij$ in six-jet events
from QCD at ${\cal O}(\alpha_s^4)$
(with $i=1, ... 5$ and $j=i+1, ... 6$). As usual in multi-jet analyses,
we first order the jets in energy, so that $E_1>E_2> ...>E_5>E_6$. Then,
we construct the quantities 
\be
\label{mij}
m_{ij}\equiv\frac{M_{ij}^2}{s}=\frac{2E_iE_j(1-\cos\theta_{ij})}{s},
\ee
where $M_{ij}^2$ represents the Lorentz invariant (squared) mass
and the equality holds for massless $ij$ particles. These fifteen quantities 
are shown in Fig.~\ref{fig_massesnlc} for the C scheme at $y=0.001$,  
their shape being similar for the D algorithm.
We found it convenient to plot the `reduced' invariant masses $m_{ij}$ rather
than the actual ones $M_{ij}$, as energies and angles `scale' with the CM 
energy in such a way that the shape of the distributions is largely unaffected
by changes of the value of $\sqrt s$ in the energy range relevant to NLC.
Therefore, from Fig.~\ref{fig_massesnlc} one should then be able to reconstruct
rather accurately the Lorentz invariant mass distributions for a given
CM energy $\sqrt s$ by exploiting eq.~(\ref{mij}). 
Note that the reduced mass spectra are similar in all three reactions 
(\ref{6j}), their integral being however rescaled according to the numbers
given in Fig.~\ref{fig_comp6nlc}. To allow for an easy conversion
of the differential cross sections 
into numbers of events in a certain mass range, 
the spectra in Fig.~\ref{fig_massesnlc} sum to the total cross section of
processes (\ref{6j}) at NLC.

It is interesting to notice in Fig.~\ref{fig_massesnlc} the `resonant'
behaviour of some of the distributions. This is particularly true
for those involving the most energetic of all the partons.
Using the definition (\ref{mij}) they translate into peak-like structures at 
the true invariant mass values $M_{ij}\approx(250)[212]\{177\}$ GeV,
for the combinations $ij=(12)[13]\{14\}$, 
and, possibly, $M_{15}\approx150$ GeV as well. In all the other cases the
spectra are generally softer and do not show any distinctive kinematic 
feature. It is however important that 
all such behaviours are correctly implemented
in the simulation programs that will be adopted by the NLC experiments, so 
that it will be possible to recognise and eventually subtract the six-jet QCD 
background, if not to study it as signal on its own in testing the fundamental
nature of QCD. Furthermore, it should be noted that the much lower value
of $\alpha_s$ at NLC energies 
(as compared to that at LEP1 and/or LEP2) in principle
implies a reduced importance of the uncalculated HO strong corrections.

In summary, we have studied the parton level processes
$e^+e^-\ar q\bar q gggg$, $e^+e^-\ar q\bar q q'\bar q' gg$ and  
$e^+e^-\ar q\bar q q'\bar q' q''\bar q''$ (for massless quarks)
at leading-order in perturbative QCD by computing their 
exact matrix elements.
The use of helicity amplitude methods has allowed the implementation
of the several hundreds of Feynman diagrams entering in such reactions
in a compact form usable for high statistic MC simulations.
It requires about $10^{-2}$ CPU seconds 
to evaluate a single event on a alpha-station DEC 3000 - M300X
and further optimisations are under way.
Some results relevant to multi-jet analyses at past, present and 
future high energy electron-positron colliders have been presented. 

\section*{Acknowledgements}

We are grateful to the UK PPARC for support. We also thank
the Theoretical Physics Group at Fermilab and that in Lund
for their kind hospitality while part of this work was carried out.
This research work is supported in part by the Italian Institute of 
Culture `C.M. Lerici' under the grant Prot. I/B1 690, 1997.

\vfill
\end{document}